\documentclass{article}
\usepackage{graphicx}
\usepackage{float}
\pdfoutput=1
\usepackage{subcaption}
\usepackage{booktabs} 
\usepackage{multirow}
\usepackage{stfloats}
\usepackage{amsmath}
\usepackage{amssymb}
\usepackage{xcolor}
\usepackage{array}
\newcommand{\PreserveBackslash}[1]{\let\temp=\\#1\let\\=\temp}
\newcolumntype{C}[1]{>{\PreserveBackslash\centering}p{#1}}
\newcolumntype{R}[1]{>{\PreserveBackslash\raggedleft}p{#1}}
\newcolumntype{L}[1]{>{\PreserveBackslash\raggedright}p{#1}}
\usepackage{caption}
\usepackage[capitalize]{cleveref}
\usepackage{cite}
\usepackage{spconf,amsmath,graphicx}

\usepackage{enumitem}
\title{Time-Frequency Attention for Monaural Speech Enhancement}
\name{Qiquan Zhang{$^{1}$}, Qi Song$^{2,*}$, Zhaoheng Ni{$^3$}, Aaron Nicolson{$^4$}, Haizhou Li$^{1,5}$\thanks{This research is supported by A*STAR under its AME Programmatic Funding Scheme (Project No. A18A2b0046) and its RIE2020 AME Programmatic Grant (Grant No. A1687b033, Project Title: Spiking Neural Networks)} \thanks{*Corresponding author}}
\address{$^1$Department of Electrical and Computer Engineering, National University of Singapore, Singapore\\
$^2$Alibaba Group, China \, $^3$Meta AI, United States \\
$^4$Australian e-Health Research Centre, CSIRO, Australia \\
$^5$Chinese University of Hong Kong, Shenzhen, China\\}

\begin{document}
\ninept
\maketitle
\vspace{-0.5em}
\begin{abstract}
Most studies on speech enhancement generally don't explicitly consider the energy distribution of speech in time-frequency (T-F) representation, which is important for accurate prediction of mask or spectra. In this paper, we present a simple yet effective T-F attention (TFA) module, where a 2-D attention map is produced to provide differentiated weights to the spectral components of T-F representation. To validate the effectiveness of our proposed TFA module, we use the residual temporal convolution network (ResTCN) as the backbone network and conduct extensive experiments on two commonly used training targets. Our experiments demonstrate that applying our TFA module significantly improves the performance in terms of five objective evaluation metrics with negligible parameter overhead. The evaluation results show that the proposed ResTCN with the TFA module (ResTCN+TFA) consistently outperforms other baselines by a large margin.

\end{abstract}
\section{Introduction}
\label{sec:intro}

Speech enhancement seeks to enhance the speech signal in the presence of background noise. It is a fundamental component for many speech processing applications, such as automatic speech recognition, speaker identification, hearing aids, and teleconference. Statistical model-based speech enhancement \cite{loizou2013,zhang2019,mmse} has been extensively studied for decades, which performs well for stationary noises, however, fails to handle non-stationary noises \cite{zhang2020}. 

Speech enhancement with supervised deep learning has achieved remarkable progress. Existing methods can be grouped into two categories by the way input signals are handled. Time-domain methods perform speech enhancement directly on the speech waveform, where a DNN is optimized to learn the mapping from the noisy waveform to the clean one \cite{SEGAN,convtasnet}. Time-frequency (T-F) domain methods typically train a DNN to predict a spectral representation of the clean speech or a T-F mask. The most popular T-F masks include ideal ratio mask (IRM) \cite{wang2014training}, phase-sensitive mask (PSM) \cite{erdogan2015phase}, and complex IRM (cIRM) \cite{williamson2015complex}. In this study, we adopt the IRM and PSM to perform speech enhancement. 


In earlier studies, multi-layer perceptrons (MLPs) are the most widely adopted architectures, but they are limited in capturing the long-term dependencies. To overcome the limitation, Chen \emph{et al.} \cite{chenlstm} employed a recurrent neural network (RNN) with four long short-term memory (LSTM) layers to perform speech enhancement, demonstrating obvious superiority over MLPs. However, LSTM network suffers from a slow and complex training procedure and requires a large number of parameters, which severely limits its applicability. Recently, the residual temporal convolution networks (ResTCNs) \cite{TCN2018}, which utilize dilated convolution and residual skip connections, have shown impressive performance in modeling long-term dependencies and gained considerable success in speech enhancement \cite{deepmmse,grn,TCNN}. More recently, self-attention based Transformer \cite{transformer} model has been successfully applied for speech enhancement and many other speech processing-related tasks for their capability of capturing long-range dependencies. 

The existing models mainly focus on how to effectively model the long-range dependencies, while they generally ignore the energy distribution characteristics of speech in T-F representation, which is equally important to speech enhancement. The attention mechanisms \cite{senet,cbam, trinh2018bubble} have been well studied to learn what is important to the learning task. Inspired by the idea of attention, we propose a novel architecture unit, termed T-F attention (TFA) module, to model the energy distribution of speech. Specifically, the TFA module consists of two parallel attention branches, i.e., time-dimension (TA) and frequency-dimension attention (FA) \cite{faa} that produce two 1-D attention maps to guide the models to focus on `where' (which time frames) and `what' (which frequency channels) respectively. The TA and FA modules are combined to generate a 2-D attention map enabling the models to capture the speech distribution in the T-F domain. To validate the idea, we use the recent ResTCN architecture as the backbone network and adopt two representative training targets, that will be discussed in Section 2,  to perform extensive experiments.


The rest of this paper is organized as follows. Section \ref{sec:2} gives the introduction of T-F domain speech enhancement. In Section \ref{sec:3}, we describe the proposed network. Section \ref{sec:4} presents the experimental setup and evaluation results. Section \ref{sec:5} concludes this paper.



\section{problem formulation}
\label{sec:2}
The noisy speech can be modeled as a combination of clean speech and additive noise in the Short-Time Fourier Transform (STFT) domain:
\begin{equation}
X[l,k]=S[l,k]+D[l,k]
\end{equation}
where $X[l,k]$, $S[l,k]$, and $D[l,k]$  denote the STFT coefficients at time frame $l$  and frequency bin $k$ of the noisy speech, clean speech, and noise, respectively. For supervised speech enhancement, a DNN is typically trained to predict the pre-designed training targets. The results are then applied to reconstruct the clean speech. To demonstrate the efficacy of our proposed TFA module, we adopt two widely used training targets to conduct extensive enhancement experiments. The details are given below.

The ideal ratio mask (IRM) \cite{wang2014training} is defined as:
\begin{equation}\label{IRM}
\text{IRM}[l,k]=\sqrt{\frac{|S[l, k]|^{2}}{|S[l, k]|^{2}+|D[l, k]|^{2}}}
\end{equation}
where $\left|S\left[l,k\right]\right|$ and $\left|D\left[l,k\right]\right|$ denote the spectral magnitudes of clean speech and noise, respectively.
The phase-sensitive mask (PSM) \cite{erdogan2015phase} is defined on the STFT magnitude of clean and noisy speech. A phase error item is introduced to compensate for utilizing the noisy speech phases:
\vspace{-1.0em}
\begin{equation}\label{PSM}
\text{PSM}[l,k]=\frac{\left|S[l, k]\right|}{|X[l, k]|}\cos[\theta_{S[l, k]-X[l, k]}]
\end{equation}
where $\theta_{S[l, k]-X[l, k]}$ denotes the phase difference between clean speech and noisy speech. The PSM is truncated to between 0 and 1 to fit the output range of the sigmoid activation function.

\section{Speech Enhancement with T-F Attention}\label{sec:3}

\subsection{Network Architecture}

\begin{figure}[!ht]
\vspace{-1.3em}
\centering
\begin{subfigure}[t]{0.4\columnwidth}
\centerline{\includegraphics[width=0.9\columnwidth]{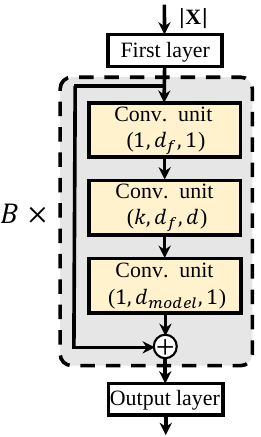}}
\caption{}
\label{fig2:1}
\end{subfigure}
\hspace{8mm}
\begin{subfigure}[t]{0.35\columnwidth}
\centerline{\includegraphics[width=0.8\columnwidth]{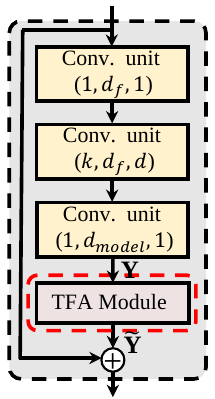}}
\caption{}
\label{fig2:2}
\end{subfigure}
\caption{Illustration of (a) the ResTCN backbone network and (b) our proposed ResTCN block with TFA module.}
\label{fig2}
\vspace{-2em}
\end{figure}

\textcolor{black}{Fig.\ref{fig2} (a) shows the architecture of the ResTCN backbone network \cite{deepmmse}, which takes as input the STFT magnitude of noisy speech $|\textbf{X}|\!\in\!\mathbb{R}^{L\times K}$ of $L$ frames, each having $K$ frequency bins. 
The output layer is a fully-connected layer with a sigmoidal activation function that generates the output masks, that is IRM or PSM. Fig.\ref{fig2}(b) shows how we plug our TFA module into the ResTCN block. The ResTCN block (shown in the black dotted box of Fig.\ref{fig2} (a)) includes three 1-D causal dilated convolution units. The kernel size, number of filters, and dilation rate for each convolutional unit are denoted as \textbf{kernel size}, \textbf{number of filters}, \textbf{dilation rate}. 
The dilation rate is cycled as the block index $b=\left\{1,2,3,...,B\right\}$ increases: $d=2^{\left(b-1 \bmod \left(\log _{2}(D)+1\right)\right.}$, where mod is the modulo operation and $D=16$ is the maximum dilation rate. Each convolutional unit employs a pre-activation design, where the input is pre-activated using frame-wise alization (LN) followed by the ReLU activation function.\\
}

\subsection{T-F Attention Module}\label{sec:3.1}
\begin{figure*}[!ht]
\vspace{-1.35em}
\begin{center}
\centerline{\includegraphics[width=1.50\columnwidth]{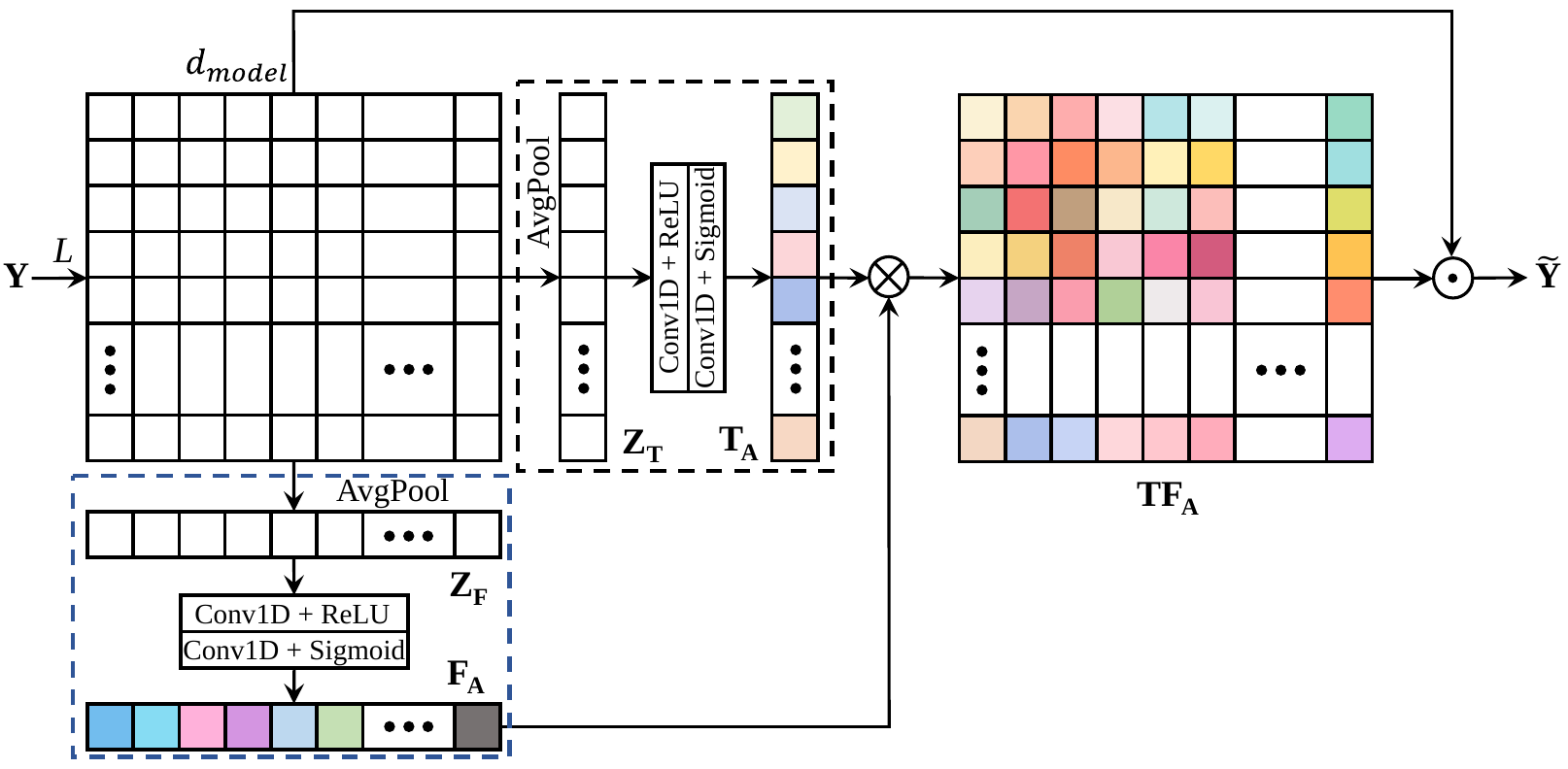}}
\vspace{-1.0em}
\caption{Diagram of our proposed TFA module, where the TA and FA modules are shown in black and blue dotted boxes, respectively. 
AvgPool and Conv1D represent average pooling and 1-D convolution operation, respectively.
$\otimes$ and $\odot$ denote the matrix multiplication and element-wise product, respectively.
}
\label{TFA}
\end{center}
\vspace{-2.5em}
\end{figure*}

In Fig. \ref{TFA}, we illustrate the proposed TFA module. We take a transformed T-F representation $\textbf{Y}\!\in \!\mathbb{R}^{L\times d_{model}}$ as input of $L$ frames and $d_{model}$ frequency channels. TFA utilizes two branches to generate a 1-D frequency-dimension attention map $\textbf{F}_{\textbf{A}}\!\in \mathbb{R}^{1 \times d_{model}}$ and a 1-D time-frame attention map $\textbf{T}_{\textbf{A}}\!\in\!\mathbb{R}^{L\times 1}$ in parallel, then combines them with a matrix multiplication to obtain the final 2-D T-F attention map $\textbf{TF}_{\textbf{A}}\!\in \mathbb{R}^{L \times d_{model}}$. The refined output is written as:
\begin{equation}
\widetilde{\textbf{Y}} = \textbf{Y} \odot \textbf{TF}_{\textbf{A}},
\end{equation}
where $\odot$ denotes the element-wise product. The detailed description of the proposed TFA attention is given below.

The energy distribution of speech along time and frequency dimension is essential for producing an accurate attention map. Each attention branch generates the attention map in two steps: global information aggregation and attention generation. Specifically, the FA module takes global average pooling along the time-frame dimension on given input $\textbf{Y}$ and generates a frequency-wise statistic $\textbf{Z}_{\textbf{F}}\in \mathbb{R}^{1 \times d_{model}}$, formulated as:
\begin{equation}
\textbf{Z}_{\textbf{F}}(k) = \frac{1}{L}\sum_{l=1}^{L} \textbf{Y}(l,k), 
\end{equation}
where $\textbf{Z}^{\textbf{F}}(k)$ is the \textit{k}-th element of $\textbf{Z}_{\textbf{F}}$. Similarly, the TA module takes global average pooling along the frequency dimension on input $\textbf{X}$ and generates a time-frame-wise statistic $\textbf{Z}_{\textbf{T}}\in \mathbb{R}^{L \times 1}$. The \textit{l}-th element of $\textbf{Z}_{\textbf{T}}$ is written as:
\begin{equation}
    \setlength{\abovedisplayskip}{5pt}
    \setlength{\belowdisplayskip}{5pt}
    \textbf{Z}_{\textbf{T}}(l) = \frac{1}{d_{model}}\sum_{k=1}^{d_{model}} \textbf{Y}(l,k). 
\end{equation}

The two statistics $\textbf{Z}_{\textbf{T}}$ and $\textbf{Z}_{\textbf{F}}$ can be seen as two descriptors for the speech energy distributions in time-frame dimension and frequency dimension, respectively. 
To make full use of the two descriptors to produce the accurate attention weights, we stack two 1-D convolution layers of size $k_{t\!f\!a}\!=\!17$ as the nonlinear transformation function. 
Specifically, the attention in the FA module is calculated as:
\begin{equation}
    \textbf{F}_{\textbf{A}} = \sigma(f_{2}^{F\!A}(\delta (f_{1}^{F\!A}(\textbf{Z}_{\textbf{F}}))))
\end{equation}
where $f$ denotes a 1-D convolution operation, $\delta$ and $\sigma$ refer to the ReLU and sigmoid activation functions, respectively. The same calculation process is applied in the TA module to generate the attention map:
\begin{equation}
\textbf{T}_{\textbf{A}} = \sigma(f_{2}^{T\!A}(\delta (f_{1}^{T\!A}(\textbf{Z}_{\textbf{T}})))).
\end{equation}
Then, the obtained attention maps from two attention branches are combined with a tensor multiplication, producing our final 2-D attention map $\textbf{TF}_{\textbf{A}}$ written as:
\begin{equation}
\textbf{TF}_{\textbf{A}} = \textbf{T}_{\textbf{A}} \otimes \textbf{F}_{\textbf{A}},
\end{equation}
where $\otimes$ denotes the tensor multiplication operation. The $(l,k)$-th element of the final 2-D attention map $\textbf{TF}_{\textbf{A}}$ is computed as:
\begin{equation}
\textbf{TF}_{\textbf{A}}(l,k) = \textbf{T}_{\textbf{A}}(l)\times \textbf{F}_{\textbf{A}}(k)
\end{equation}
where $\textbf{T}_{\textbf{A}}(l)$ and $\textbf{F}_{\textbf{A}}(k)$ denote the $l$-th element of $\textbf{T}_{\textbf{A}}$ and the $k$-th element of $\textbf{F}_{\textbf{A}}$, respectively.

\section{EXPERIMENTS}
\label{sec:4}
\subsection{Datasets and Feature Extraction}
We use the \textit{train-clean-100} set from the Librispeech  \cite{panayotov2015librispeech} corpus as the clean speech recordings in the training set, which includes $28\,539$ utterances spoken by $251$ speakers. The employed noise recordings in the training set are taken from the following datasets: the QUT-NOISE dataset \cite{dean2010qut}, the Nonspeech dataset \cite{hu2004100}, the Environmental Background Noise dataset \cite{saki2016smartphone,saki2016automatic}, the RSG-10 dataset \cite{steeneken1988description} (\textit{voice babble}, \textit{F16}, and \textit{factory welding} are excluded for testing), the Urban Sound dataset \cite{Urban} (\textit{street music} recording no.26\,270 is excluded for testing), the noise set from the MUSAN corpus \cite{snyder2015musan}, and coloured noise recordings (with an $\alpha$ value ranging from $-2$ to $2$ in increments of 0.25). This gives a total $6\,909$ noise recordings. For the validation set, we randomly select $1\,000$ clean speech and noise recordings (without replacement) and remove them from the aforementioned clean speech and noise sets. 
Each clean speech is mixed with a random section of one noise recording at a random SNR level between -10 dB and 20 dB in 1 dB increments, which generates $1\,000$ noisy speech as the validation set. For the test set we utilize the four real-world noise recordings (\textit{voice babble}, \textit{F16}, \textit{factory welding}, and \textit{street music}) excluded from the RSG-10 dataset \cite{steeneken1988description} and Urban Sound dataset \cite{Urban}. 
For each of the four noise recordings, ten clean speech recordings (without replacement) randomly selected from the \textit{test-clean-100} of Librispeech corpus \cite{panayotov2015librispeech} are mixed with a random segment of the noise recordings at the following SNR levels: \{-5 \text{dB}, 0 \text{dB}, 5 \text{dB}, 10 \text{dB}, 15 \text{dB}\}. This generates a test set of 200 noisy speech recordings. All clean speech and noise recordings are single-channel, with a sampling frequency of 16 kHz.

A square-root-Hann window function is used for spectral analysis and synthesis, with a frame-length of 32 ms and a frame-shift of 16 ms. The 257-point single-sided STFT magnitude spectrum of noisy speech, which includes both the DC frequency component and the Nyquist frequency component is used as the input.
\subsection{Experimental Setup}
The ResTCN model is used as the baseline backbone to validate the effectiveness of our TFA module. In addition, we also adopt two recent models as baselines, the ResTCN with self attention (ResTCN+SA) \cite{restcnsa} and the multi-head self-attention network (MHANet) \cite{mhanet}. The ResTCN baseline employs the following the parameter as in \cite{deepmmse}, $k\!=\!3$, $d_{{model}}\!=\!256$, $d_{f}\!=\!64$, and $B\!=\!40$. 
ResTCN+SA \cite{restcnsa} employs a multi-head self-attention module to produce dynamic representations followed by a ResTCN model ($B\!=\!40$ stacked baseline ResTCN blocks are adopted to build the ResTCN model for a fair comparison) to perform nonlinear mapping. 
MHANet model \cite{mhanet} uses $5$ stacked Transformer encoder layers \cite{transformer} to perform speech enhancement, with the parameter setting as in \cite{mhanet}. To validate the efficacy of FA and TA components in TFA module, we conduct an ablation study, where ResTCN using FA and TA (termed ResTCN+FA and ResTCN+TA) are evaluated.

\textbf{Training methodology}: A mini-batch size of $10$ noisy speech utterances is used for each training iteration. The noisy speech signals are created as follows: each clean speech recording selected for the mini-batch is mixed with a random section of a randomly selected noise recording at a randomly selected SNR level (-10 dB to 20 dB, in 1 dB increments). The mean squared error (MSE) between the target mask and the estimated mask is used as the objective function. For ResTCN, ResTCN+SA, and proposed models, the \textit{Adam} optimizer with default hyper-parameters \cite{Adam} and a learning rate of $0.001$ are used for gradient descent optimisation. As MHANet has been found difficult to train \cite{mhanet,difficulty}, we employ the training strategy as in \cite{mhanet}. Gradient clipping is applied to all the models, where the gradients are clipped between $[-1,1]$. 
\vspace{-1.0em}

\subsection{Training \& Validation Error}

\begin{figure}[!hbpt]
\vspace{-0.8em}
\centering
\begin{subfigure}[t]{0.495\columnwidth}
\centerline{\includegraphics[width=\columnwidth]{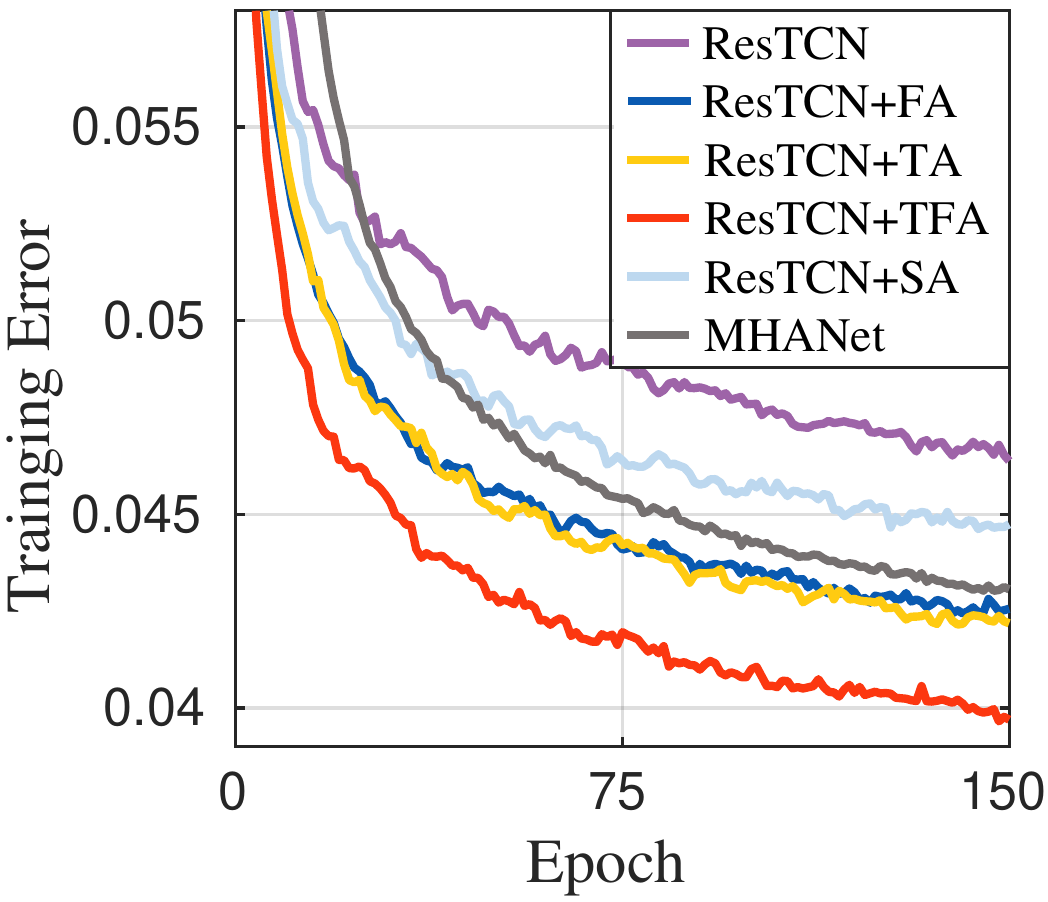}}
\caption{}
\label{fig3:1}
\end{subfigure}
\begin{subfigure}[t]{0.495\columnwidth}
\centerline{\includegraphics[width=\columnwidth]{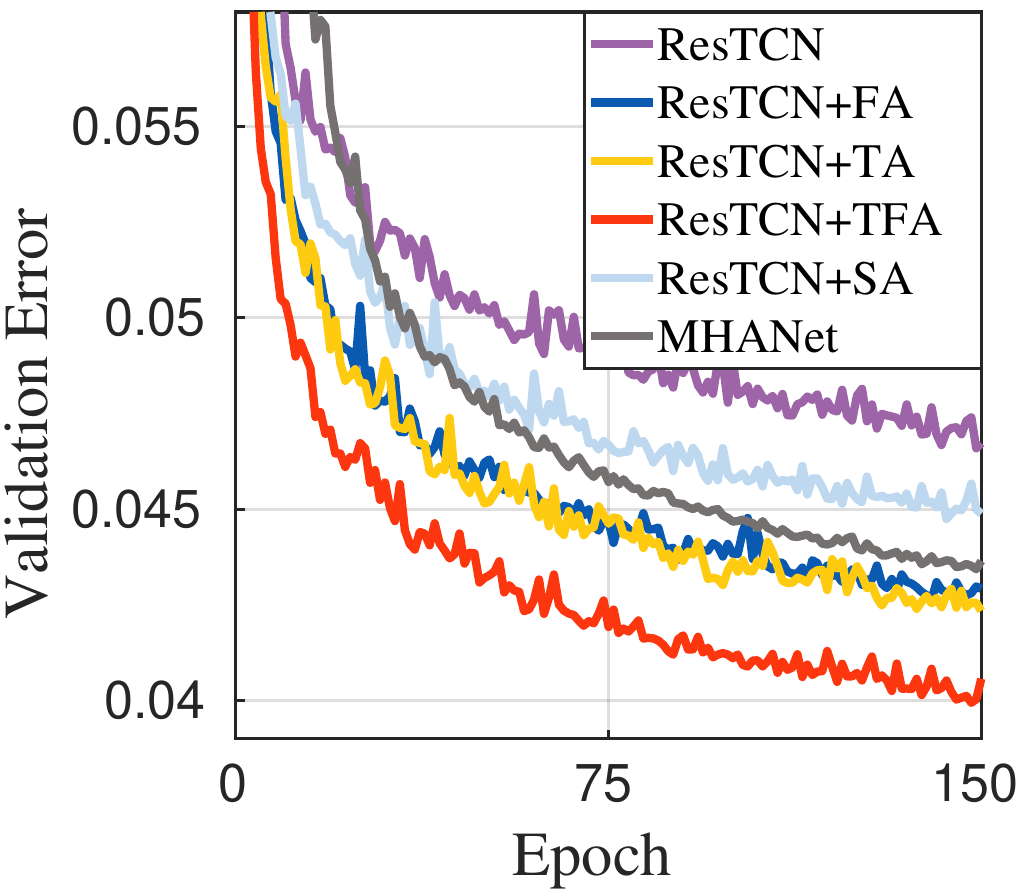}}
\caption{}
\label{fig3:2}
\end{subfigure}
\vspace{-1.0em}
\caption{The curves of training error (a) and validation error (b) on the IRM training target.}
\label{fig3}
\vspace{-1em}
\end{figure}

\begin{figure}[!ht]
\vspace{-0.5em}
\centering
\begin{subfigure}[t]{0.495\columnwidth}
\centerline{\includegraphics[width=\columnwidth]{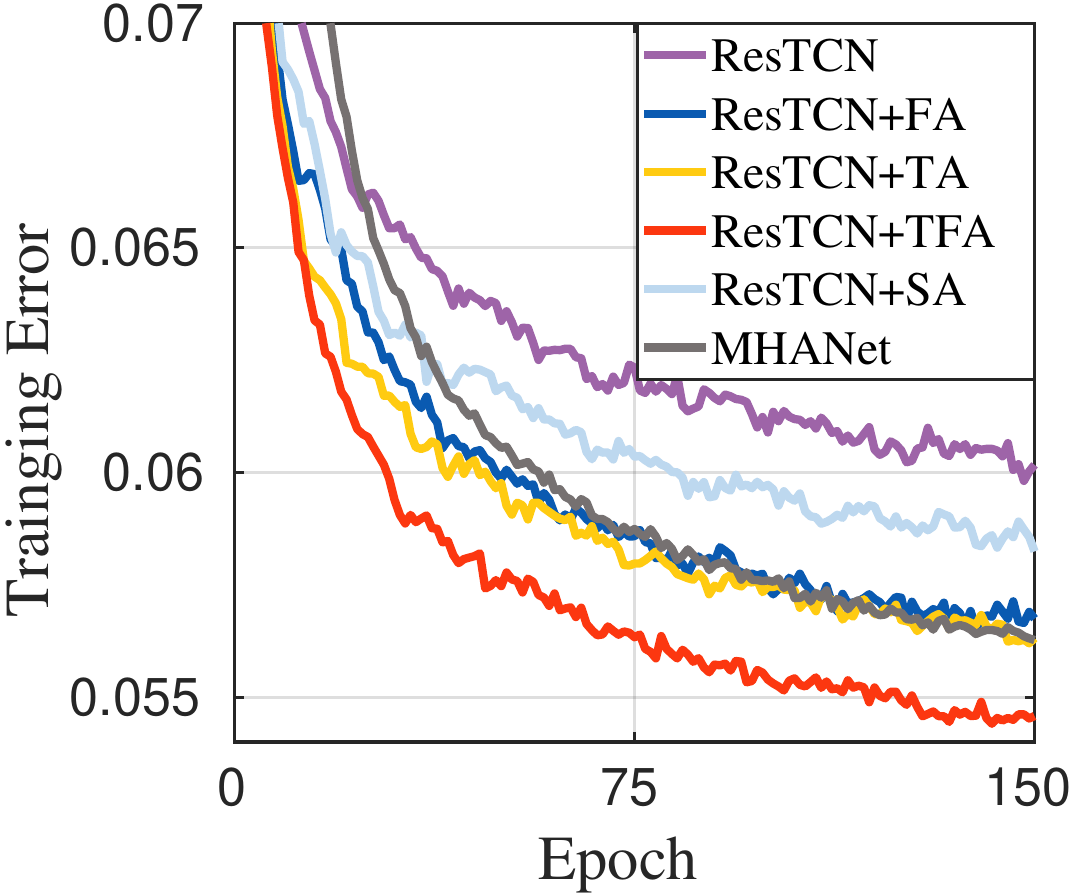}}
\caption{}
\label{fig4:1}
\end{subfigure}
\begin{subfigure}[t]{0.495\columnwidth}
\centerline{\includegraphics[width=\columnwidth]{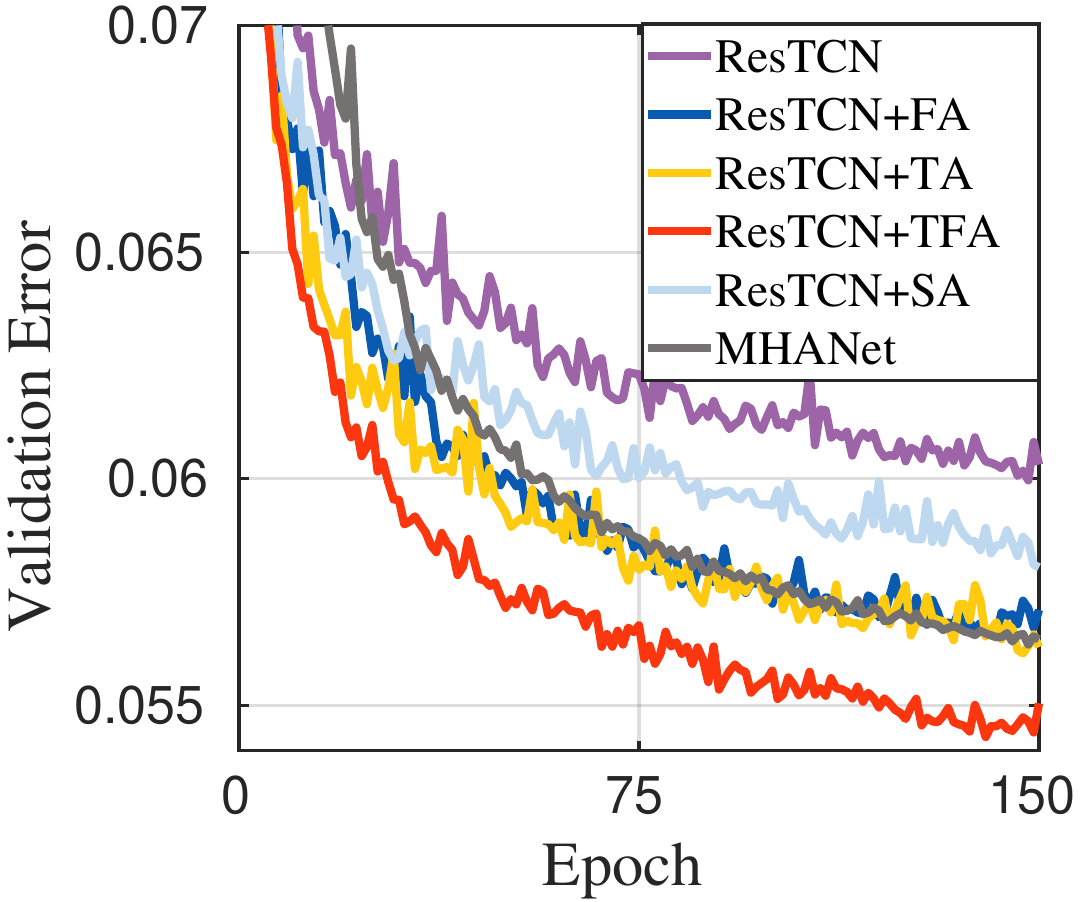}}
\caption{}
\label{fig4:2}
\end{subfigure}
\vspace{-1.0em}
\caption{The curves of training error (a) and validation error (b) on the PSM training target.}
\label{fig4}
\vspace{-1.5em}
\end{figure}

Fig. \ref{fig3}-\ref{fig4} show the curves of training and validation errors produced by
each of the models for 150-epoch training. It can be seen that the ResTCN with our proposed
TFA (ResTCN+TFA) yields significantly lower training and validation errors as compared to ResTCN, which confirms the efficacy of the TFA module. Meanwhile, compared to ResTCN+SA and MHANet, ResTCN+TFA achieves the lowest training and validation errors, and shows obvious superiority. Among the three baselines, MHANet performs best, and ResTCN+SA outperforms ResTCN. In addition, the comparisons among ResTCN, ResTCN+FA, and ResTCN+TA demonstrate the efficacy of TA and FA modules. 

\subsection{Results and Discussion}
In this study, five metrics are used for extensive evaluations of enhancement performance, including wideband perceptual evaluation of speech quality (PESQ) \cite{pesq}, extended short-time objective intelligibility (ESTOI) \cite{jensen2016algorithm}, and three composite metrics \cite{hu2007evaluation}, which are mean opinion score (MOS) predictors of the signal distortion (CSIG), background-noise intrusiveness (CBAK), and overall signal quality (COVL). 

\begin{table}[!h]
    \vspace{-1.0em}
    \centering
    \scriptsize
    \def\arraystretch{1.2}
    \setlength{\tabcolsep}{4.5pt}
    \footnotesize	
    \caption{Average PESQ (wideband version) scores for each SNR level. The highest PESQ scores are highlighted in boldface.}
    \label{tabwer}
    \vspace*{0.5em}
    \scalebox{0.92}{\begin{tabular}{clrccccc}
        \toprule
        &  &  & \multicolumn{5}{c}{\textbf{Input SNR (dB)}}\\  
        \cmidrule[0.8pt]{4-8}
        {\textbf{Target}} & {\textbf{Network}} & {\textbf{\# Params}} & -5 & 0 & 5 & 10 & 15 \\
        \toprule
        - & Noisy & & 1.05 & 1.07 & 1.13 & 1.31 & 1.64 \\
        \toprule
        \multirow{5}{*}{\textbf{IRM}}
        & ResTCN+SA \cite{restcnsa} & 2.24M & 1.15 & 1.34 & 1.64 & 2.07 & 2.51  \\
        & MHANet \cite{mhanet}      & 4.08M & 1.16 & 1.36 & 1.68 & 2.10 & 2.56  \\
        \cmidrule[0.8pt]{2-8}
        & ResTCN 
        \cite{deepmmse} & 1.98M  & 1.13 & 1.32 & 1.61 & 2.06 & 2.50  \\
        & ResTCN+FA     & +1.36K & 1.18 & 1.39 & 1.74 & 2.15 & 2.60 \\
        & ResTCN+TA     & +1.36K & 1.18 & 1.40 & 1.75 & 2.19 & 2.63 \\
        & ResTCN+TFA    & +2.72K & \textbf{1.20} & \textbf{1.43} & \textbf{1.79} & \textbf{2.27} & \textbf{2.70}\\
        \toprule
        \toprule
        
        \multirow{5}{*}{\textbf{PSM}}
        & ResTCN+SA \cite{restcnsa} & 2.24M & 1.20 & 1.42 & 1.79 & 2.26 & 2.75  \\
        & MHANet \cite{mhanet}      & 4.08M & 1.20 & 1.44 & 1.83 & 2.30 & 2.78  \\
        \cmidrule[0.8pt]{2-8}
        & ResTCN \cite{deepmmse}    & 1.98M & 1.19 & 1.40 & 1.76 & 2.23 & 2.72  \\
        & ResTCN+FA   & +1.36K & 1.20 & 1.47 & 1.87 & 2.36 & 2.85\\
        & ResTCN+TA   & +1.36K & 1.21 & 1.49 & 1.90 & 2.38 & 2.85\\
        & ResTCN+TFA   & +2.72K & \textbf{1.26} & \textbf{1.54} & \textbf{1.96} & \textbf{2.44} & \textbf{2.91}\\
        \bottomrule
    \end{tabular}}
    \label{pesq}
\end{table}
\begin{table}[!htbp]
    \centering
    \scriptsize
    \def\arraystretch{1.2}
    \setlength{\tabcolsep}{4.5pt}
    \footnotesize	
    \caption{Average ESTOI scores (in \%) for each SNR level. The highest ESTOI scores are highlighted in boldface.}
    \label{tabwer}
    \vspace{0.5em}
    \scalebox{0.90}{\begin{tabular}{clrccccc}
        \toprule
        & & & \multicolumn{5}{c}{\textbf{Input SNR (dB)}}\\  
        \cmidrule[0.8pt]{4-8}
        \textbf{Target} & \textbf{Network} & {\textbf{\# Params}} & -5 & 0 & 5 & 10 & 15 \\
        \toprule
        - & Noisy & & 27.91 & 42.14 & 57.21 & 71.11 & 82.22 \\
        \toprule
        \multirow{5}{*}{\textbf{IRM}}
        & ResTCN+SA \cite{restcnsa}    & 2.24M & 43.32 & 60.68 & 74.37 & 83.67 & 89.77  \\
        & MHANet \cite{mhanet}       & 4.08M & 43.72 & 61.08 & 74.68 & 83.90 & 89.96  \\
        \cmidrule[0.8pt]{2-8}
        & ResTCN \cite{deepmmse}       & 1.98M  & 42.47 & 59.93 & 73.67 & 83.22 & 89.44  \\
        & ResTCN+FA    & +1.36K & 44.87 & 61.73 & 75.06 & 84.27 & 90.09\\
        & ResTCN+TA    & +1.36K & 46.11 & 62.96 & 75.86 & 84.48 & 90.21\\
        & ResTCN+TFA   & +2.72K & \textbf{47.61} & \textbf{64.05} & \textbf{78.61} & \textbf{85.45} & \textbf{90.83}\\
        \toprule
        \toprule
        
        \multirow{5}{*}{\textbf{PSM}}
        & ResTCN+SA \cite{restcnsa}    & 2.24M & 43.03 & 61.13 & 75.07 & 84.24 & 90.07  \\
        & MHANet \cite{mhanet}       & 4.08M & 44.66 & 62.32 & 75.74 & 84.69 & 90.43  \\
        \cmidrule[0.8pt]{2-8}
        & ResTCN \cite{deepmmse}       & 1.98M  & 42.76 & 60.45 & 74.47 & 83.91 & 89.84  \\
        & ResTCN+FA    & +1.36K & 44.91 & 62.38 & 75.88 & 84.82 & 90.45\\
        & ResTCN+TA    & +1.36K & 45.70 & 63.25 & 76.49 & 85.09 & 90.70\\
        & ResTCN+TFA   & +2.72K & \textbf{48.67} & \textbf{65.28} & \textbf{77.43} & \textbf{85.80} & \textbf{91.16}\\

        \bottomrule
    \end{tabular}}
    \label{estoi}
\end{table}
Tables \ref{pesq} and \ref{estoi} present the average PESQ and ESTOI scores for each SNR level (across four noise sources), respectively. The evaluation results show that our proposed ResTCN+TFA consistently achieves significant improvements over ResTCN in terms of PESQ and ESTOI on IRM and PSM, with negligible parameter overhead, which demonstrates the effectiveness of the TFA module. In the 5 dB SNR case, for instance, ResTCN+TFA with IRM improves the baseline ResTCN by 0.18 in PESQ and by 4.94\% in ESTOI. Compared to MHANet and ResTCN+SA, ResTCN+TFA performs best in all cases, and shows obvious performance superiority. Among the three baselines, overall, the performance rank is in order MHANet $\textgreater$ ResTCN+SA $\textgreater$ ResTCN. 
Meanwhile, ResTCN+FA and ResTCN+TA also provide substantial improvements over ResTCN, which confirms the efficacy of FA and TA modules. 
Table \ref{composite} lists the average CSIG, CBAK, and COVL scores across all of the test conditions. Similar performance trends are observed to those in Tables \ref{pesq} and \ref{estoi}. Again, our proposed ResTCN+TFA significantly outperforms ResTCN in three metrics, and performs best among all models. On average, ResTCN+TFA with PSM improves CSIG by 0.21, CBAK by 0.12, and COVL by 0.18 over ResTCN. Compared to MHANet, ResTCN+TFA with PSM improves CSIG by 0.12, CBAK by 0.08, and COVL by 0.11.

\renewcommand\arraystretch{1.0}
\vspace{-1.0em}
\begin{table}[!hbpt]
\centering
    \scriptsize
    \def\arraystretch{1.2}
    \setlength{\tabcolsep}{2.3pt}
\caption{Average scores of CSIG, CBAK, and COVL across all test conditions. The highest scores are highlighted in boldface.}
\label{tabwer}
\vspace*{0.1in}
\scalebox{1}{
\begin{tabular}{l|ccc|ccc }
\toprule
\multirow{2}{*}{\textbf{Network}}  & \multicolumn{3}{c|}{\textbf{IRM}} & \multicolumn{3}{c}{\textbf{PSM}}  \\
\cline{2-7}
 &  \textbf{CSIG} & \textbf{CBAK} & \textbf{COVL} & \textbf{CSIG} & \textbf{CBAK} & \textbf{COVL} \\
\toprule 
Noisy & 2.26  & 1.80  & 1.67  & - & - & -  \\
\toprule   
ResTCN+SA \cite{restcnsa}     
& 3.10 & 2.45 & 2.37 & 3.14 & 2.53 & 2.46 \\
MHANet \cite{mhanet}      
& 3.13 & 2.46 & 2.40 & 3.21 & 2.56 & 2.51 \\
\toprule 
ResTCN \cite{deepmmse}     
& 3.08 & 2.43 & 2.35 & 3.12 & 2.52 & 2.44 \\
ResTCN+FA    
& 3.21 & 2.51 & 2.48 & 3.25 & 2.61 & 2.55 \\      
ResTCN+TA       
& 3.23 & 2.53 & 2.49 & 3.27 & 2.58 & 2.57 \\
ResTCN+TFA    
& \textbf{3.28} &\textbf{2.56} & \textbf{2.54} & \textbf{3.33} & \textbf{2.64} & \textbf{2.62} \\  
\toprule  
\end{tabular}}
\label{composite}
\vspace{-2em}
\end{table}


\section{CONCLUSION}
\label{sec:5}
In this study, we propose a light-weight and flexible attention unit termed TFA module, which is designed to model the energy distribution of speech in the T-F representation. Extensive experiments with ResTCN as a backbone on two training targets (IRM and PSM) demonstrate the effectiveness of the proposed TFA module. Among all the models, our proposed ResTCN+TFA consistently performs best and significantly outperforms other baselines in all of the cases. Future research work includes investigating the efficacy of TFA on more architectures (e.g., more recent Transformer) and more training targets.

\vfill\pagebreak

\bibliographystyle{IEEEtran}
{\linespread{0.8}
   \bibliography{refs}
   }

\end{document}